\documentclass[aps,prd,reprint,preprintnumbers,superscriptaddress,showpacs,twocolumn]{revtex4-1}
\usepackage{latexsym,graphicx,amssymb,amsmath,mathrsfs}
\usepackage{setspace,bm}
\usepackage[breaklinks, colorlinks=true, pdfstartview=FitV, linkcolor=red, citecolor=blue, urlcolor=blue]{hyperref}

\usepackage{amsmath}
\usepackage[usenames]{color}
\usepackage{latexsym}
\usepackage{epstopdf}

\newcommand\nc{N_\mathrm{c}}
\newcommand\nf{N_\mathrm{f}}
\newcommand\zc{\mathbb{Z}_{N_\mathrm{c}}}
\newcommand\mur{\mu_\mathrm{R}}
\newcommand\mui{\mu_\mathrm{I}}
\newcommand\trw{T_\mathrm{RW}}
\newcommand\qrw{\theta_\mathrm{RW}}
\usepackage[normalem]{ulem}  % \sout{old text} for strikeout

\newcommand{\comment}[1]{}

\renewcommand\sout{\bgroup \color{red} \ULdepth=-.5ex \ULset}
\begin{document}
\preprint{YITP-16-13}

\title{Quark number holonomy and confinement-deconfinement transition}

\author{Kouji Kashiwa}
\email[]{kouji.kashiwa@yukawa.kyoto-u.ac.jp}
\affiliation{Yukawa Institute for Theoretical Physics,
Kyoto University, Kyoto 606-8502, Japan}

\author{Akira Ohnishi}
\email[]{ohnishi@yukawa.kyoto-u.ac.jp}
\affiliation{Yukawa Institute for Theoretical Physics,
Kyoto University, Kyoto 606-8502, Japan}

\begin{abstract}
We propose a new quantity which describes the
 confinement-deconfinement transition based on topological
 properties of QCD.
The quantity which we call the quark number holonomy is defined as
 the integral of the quark number susceptibility along the closed loop
 of $\theta$ where $\theta$ is the dimensionless
 imaginary chemical potential.
Expected behavior of the quark number holonomy at finite temperature is
 discussed and its asymptotic behaviors are shown.
\end{abstract}

\pacs{11.30.Rd, 21.65.Qr, 25.75.Nq}
\maketitle

\section{Introduction}

Understanding the confinement-deconfinement transition in quantum
chromodynamics (QCD) is one of the interesting and important subjects in
nuclear and elementary particle physics.
In the heavy quark mass limit, spontaneous breaking of the center
($\zc$) symmetry is directly related to the confinement-deconfinement
transition, where $\nc$ is the number of color.
Then, the holonomy which is the gauge invariant integral along the
closed Euclidean temporal coordinate loop becomes an exact
order-parameter of the confinement-deconfinement transition.
It is so called the Polyakov-loop.
On the other hand, we cannot find any exact order-parameters in the case
with dynamical quarks at present, where the direct relation between
$\zc$ symmetry and the confinement-deconfinement transition is lost.
%(The following is shifted to later.)

{\em Topological order ---}
The notion of the topological order may be of great help
in understanding the nature of the confinement-deconfinement transition.
Recently, there is an important progress that the confined and
deconfined states at zero temperature ($T=0$) are mathematically classified based on the
topological order \cite{Wen:1989iv} in Ref.~\cite{Sato:2007xc}.
Motivated by the progress, it has been suggested that the
confinement-deconfinement transition can be described by using the
analogy of the topological order and then the free-energy degeneracy
plays a crucial role~\cite{Kashiwa:2015tna}.
The idea of the topological order is extended to finite temperature
in terms of the Uhlmann phase~\cite{Uhlmann1986,Viyuela2014}.
The Uhlmann phase is an extension of the Berry phase to mixed quantum
states.
The Uhlmann phase can describe the topological
order at finite $T$ in the one-dimensional fermion systems such as the
topological insulator and the superconductor~\cite{Viyuela2014}.
The Uhlmann phase is defined by using the amplitude for the density
matrix where amplitudes form the Hilbert space.
There is the $U(n)$ gauge freedom of the amplitude where $n$ is the
dimension of the space and it is a generalization of the $U(1)$ gauge
freedom of pure quantum states.
At finite $T$, the Uhlmann phase includes information of the density
matrix of the statistical mechanics and is calculated by the contour
integral along the crystalline momentum.
Unfortunately, the calculation of the Uhlmann phase in QCD seems to be
very difficult or impossible at present,

{\em Imaginary chemical potential ---}
In QCD at finite $T$, imaginary chemical potential ($\mui$) is an
external parameter, which shows periodicity;
Chemical potential $\mu$ appears in the form of fugacity
$e^{\mu/T}=e^{\mur/T}e^{i\mui/T}$
in the free-energy, and the two states at $\mui=0$ and $\mui=2\pi T$
are physically the same. In addition to this $2\pi T$ periodicity,
characteristic periodicity appears at finite
imaginary chemical potential ($\mui$).
It is so called the Roberge-Weiss (RW) periodicity~\cite{Roberge:1986mm}.
The RW periodicity has deep relations with the free-energy degeneracy
and thus it is natural to expect that some hints to understand the
confinement-deconfinement transition are hidden in the imaginary
chemical potential region.

{\em Quark number holonomy ---}
We investigate the confinement-deconfinement transition
by using the imaginary chemical potential in this paper.
We discuss the contour integral along the closed loop
of the imaginary chemical potential, $\mui=0 \sim 2\pi T$, or the
dimensionless quark imaginary chemical potential, $\theta=\mui/T=0 \sim
2\pi$.
Particularly, we focus on the behavior of the quark number density at
finite $\mui$ and propose a new quantity which can describe the
confinement-deconfinement transition based on it.
It is a new quantum order-parameter of the
confinement-deconfinement transition when dynamical quarks are acting
in the system.
We call it the {\it quark number holonomy}.
The quark number holonomy seems to be a similar quantity with the
Uhlmann phase~\cite{Uhlmann1986,Viyuela2014}.
The quark number holonomy defined in Eq.~(\ref{Eq:holonomy_n}) also
includes the information of the density matrix via the quark number
density and is calculated by the contour integral along the closed
loop of $\theta$.
It should be noted that the quark number holonomy can be calculated in
the effective models of QCD and lattice QCD simulation as discussed
later.
It is the most important reason why we propose the new quantity for the
confinement-deconfinement transition in this paper.

This paper is organized as following.
In the next section, we propose new quantity which is so called the
quark number holonomy.
Some discussions for the quark number holonomy are shown in
Sec.~\ref{Sec:Discussion}.
Section \ref{Sec:Summary} is devoted to summary.

\section{Quark number holonomy}

In this section, we firstly summarize QCD periodicities and
special transitions which appear at finite $\mui$.
Secondly, we propose a new quantity which describes the
confinement-deconfinement transition based on QCD properties at finite
$\mui$.
Finally, the infinite $T$ and the infinite bare quark mass ($m$) limit
are discussed.

\subsection{QCD periodicities and transitions at finite $\mui$}

It is known that the QCD partition function
($Z_\mathrm{QCD}$) has the RW periodicity~\cite{Roberge:1986mm};
\begin{align}
Z_\mathrm{QCD} (\theta)
&= Z_\mathrm{QCD}
   \Bigl( \theta + \frac{2\pi k}{\nc} \Bigr),
\end{align}
where $k$ is any integer.
It should be noted that the RW periodicity is a model independent
and exact property of the QCD partition function.
In the pure gauge limit, there is the $\zc$ symmetry, but it is
explicitly broken by dynamical quark contributions.
The RW periodicity is nothing but the remnant of the $\zc$ symmetry in
the pure gauge limit.
If we neglect the $\theta$ dependence of the gauge field through quark
contributions, the RW periodicity is lost and then the partition
function only has the trivial $2\pi$ periodicity.
This situation also appears in the quenched approximation.

In addition to the RW periodicity, QCD has special transition at
$\theta=(2k-1)\pi/\nc$ which is called the RW transition.
The RW periodicity is realized in a different way at the RW transition
in the confined and deconfined phases.
To discuss the RW transition, the phase of the Polyakov-loop is useful.
The Polyakov-loop can be expressed as
\begin{align}
\Phi
&= \frac{1}{\nc} \mathrm{tr} {\cal P}
   \Bigl[
   \exp \Bigl( i g \oint_0^{\beta} A_4(\tau,{\vec x})~d\tau \Bigr)
   \Bigr]
      = |\Phi| \hspace{0.5mm} e^{i \phi},
\end{align}
where $\beta$ is the inverse temperature ($\beta=1/T$),
${\cal P}$ is the path-ordering operator and $\phi$ is the
Polyakov-loop phase.
When $\theta$ is continuously changed from $0$ to $2\pi$, the phase of
the Polyakov-loop is smoothly rotated below $\trw$, but it becomes
discontinuous above $\trw$ at $\theta=(2k-1)\pi/\nc$.
Such $\theta$ dependence of $\phi$ can be found in
Ref.~\cite{Sakai:2009dv} for
the Polyakov-loop extended Nambu--Jona-Lasinio (PNJL)
model~\cite{Fukushima:2003fw}
and Ref.~\cite{deForcrand:2002hgr,D'Elia:2002gd,Wu:2006su} for lattice QCD.
The endpoint of the RW transition is called the RW endpoint and its
temperature is denoted by $\trw$.

\subsection{Deconfinement transition from RW periodicity}

In Ref. \cite{Kashiwa:2015tna}, the authors proposed the new
classification of the confined and deconfined phases at finite $T$ based
on the RW periodicity.
The different realization of the RW periodicity plays a crucial role in
the classification:
\begin{description}
 \item[Confined phase]
 The origin of the RW periodicity is the dimensionless baryon chemical
	    potential $3 \theta$ in the form of $\exp(\pm 3 i \theta)$.
 For example, it can be seen from the strong coupling limit of QCD with
	    the mean-field approximation;
 see Ref.~\cite{Nishida:2003fb,Kawamoto:2005mq}.
\item[Deconfined phase]
 The origin is the dimensionless quark chemical potential and the gauge
	   field in the form of $\exp[\pm i (gA_4/T + \theta)]$ where
	   $\mathbb{Z}_3$ images are important~\cite{Roberge:1986mm}.
 It can be clearly seen in the perturbative one-loop effective
	   potential~\cite{Gross:1981br,Weiss:1980rj}.
\end{description}
Therefore, in our approach for the investigation of the
confinement-deconfinement transition,
we focus on the response of the system against
$\theta$ as an indicator of the non-trivial free-energy degeneracy.
The system does not show singularities along $\theta$ at ($T$, $\mur$)
in the confined phase.
By comparison, there should be some singularities along $\theta$ at
($T$, $\mur$) in the deconfined phase.
Details of singularities are explained in Sec.~\ref{Sec:DQNH};
for example, see Fig.~\ref{Fig:QND}.
Confinement-deconfinement transition temperatures determined by
the non-trivial free-energy degeneracy and the Polyakov-loop are matched
with each other in the infinite quark mass limit.
In the next subsection, we propose a new quantum order parameter of the
confinement-deconfinement transition based on the RW periodicity.
In the following discussions in this section, we concentrate on the case
with $\mur=0$.

\subsection{Definition of quark number holonomy}
\label{Sec:DQNH}

\begin{figure}[t]%[H]
\begin{center}
 \includegraphics[width=0.4\textwidth]{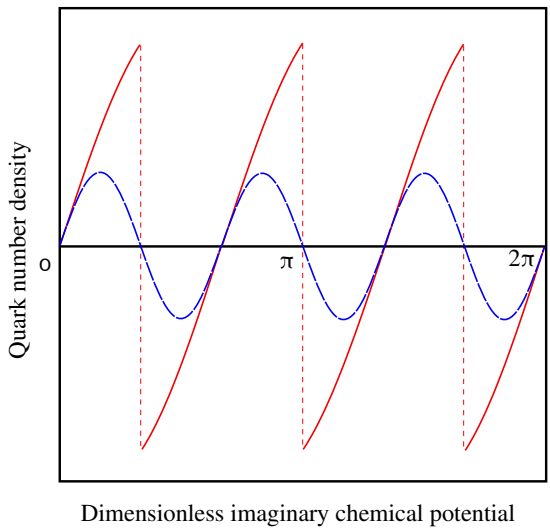}
\end{center}
\caption{
The schematic behavior of $n_q$ as a function of
 $\theta$ for $\nc=3$.
The periodic solid and dashed lines represent the quark number density
 at $T \gg T_\mathrm{RW}$ and $T \ll T_\mathrm{RW}$, respectively.
}
\label{Fig:QND}
\end{figure}

\begin{figure}[b]%[H]
\begin{center}
 \includegraphics[width=0.4\textwidth]{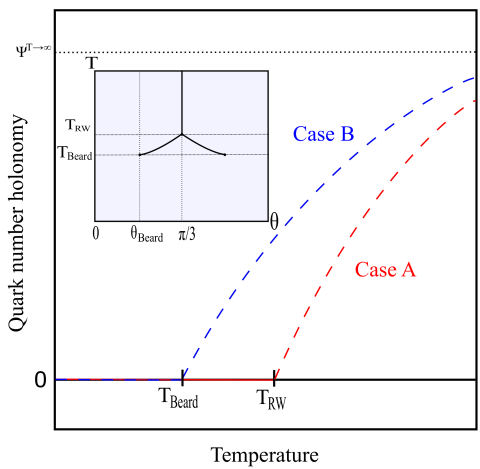}
\end{center}
\caption{
The expected behavior of $\Psi$ as a function of $T$.
In the case A, the RW endpoint is the second order,
while it is the triple point in the case B.
The actual value of $\Psi^{T \to \infty}$ is explained in the text of
Sec.~\ref{Sec:AB}.
The inset figure shows the schematic phase diagram in the case B
with $\nc=3$ as a function of $\theta$ and $T$.
}
\label{Fig:QNH}
\end{figure}

The quark number density ($n_q$) above $\trw$ should have the gap at
$\theta=(2k-1)\pi/\nc$ which reflects the $\theta$-odd property.
The schematic behavior of $n_q$ with $\nc=3$
is shown in Fig.~\ref{Fig:QND}.
The periodic solid and dashed lines represents $n_q$ at
sufficiently high and low $T$ comparing with $\trw$, respectively.
%%%%%%%%%%%%%%%%%%%% Fig %%%%%%%%%%%%%%%%%%%%%%%%
%\begin{figure}[htbp]%[H]
%%%%%%%%%%%%%%%%%%%%%%%%%%%%%%%%%%%%%%%%%%%%%%%%%
By using the behavior of $n_q$ at finite $T$, we can construct
the order-parameter;
\begin{align}
\Psi (T)
&= \Bigl[
   \oint_{0}^{2\pi}
   \Bigl\{\mathrm{Im} \Bigl(
          \frac{d {\tilde n}_q}{d \theta} \Bigl|_T \Bigr) \Bigr\}
         ~d\theta
   \Bigr],
\label{Eq:holonomy_n}
\end{align}
where ${\tilde n}_q$ is the normalized quark number density defined as
${\tilde n}_q \equiv C n_q$ here
the coefficient $C$ [MeV$^{-3}$] is introduced to make ${\tilde n}_q$
dimensionless.
It becomes non-zero at $T \gg \trw$ and zero at $T \ll \trw$ because the
information of the gap at $\theta=(2k-1)\pi/\nc$ is missed when we
perform the differential calculus and the numerical integration.
We call Eq.~(\ref{Eq:holonomy_n}) the {\it quark number holonomy}.
The integrand of Eq.~(\ref{Eq:holonomy_n}) can be expressed as
\begin{align}
\mathrm{Im} \Bigl( \frac{d {\tilde n}_q}{d \theta} \Bigl|_T \Bigr)
&= - \frac{C T^2}{V}
     \Bigl[
            \langle {\tilde N} \rangle^2 - \langle {\tilde N}^2 \rangle
     \Bigr] \Bigl|_T
 \propto \chi_2^q,
\label{Eq:qns}
\end{align}
where
$V$ denotes the three-dimensional
volume and the operator ${\tilde N}$ is $\int(q^\dag q) d^4 x$.
In Eq.~(\ref{Eq:qns}), $\chi^q_2$ is nothing but the quark
number susceptibility at finite $\theta$.
The expected behavior of the quark number holonomy as a function of $T$
is shown in Fig.~\ref{Fig:QNH}.
We assume that the RW endpoint is the second (first) order in the case A
(B).
The schematic phase diagram in the case B is shown in the inset
figure of Fig.~\ref{Fig:QNH}.
%%%%%%%%%%%%%%%%%%%% Fig %%%%%%%%%%%%%%%%%%%%%%%%
%\begin{figure}[htbp]%[H]
%%%%%%%%%%%%%%%%%%%%%%%%%%%%%%%%%%%%%%%%%%%%%%%%%
When $\trw$ is the first-order, the RW endpoint can have
two more first-order lines.
In this paper, we call it beard line and
the endpoint temperature of the beard line is denoted by
$T_\mathrm{Beard}$.
This triple point scenario has been predicted by the lattice QCD
simulations~\cite{D'Elia:2009qz,Bonati:2010gi}.
This behavior may be induced by the correlation between the chiral and
deconfinement dynamics, but details are still under debate.

In the case A where the RW endpoint is the second order, the quark
number holonomy can be expressed as
\begin{align}
\Psi
&= \pm
   2 \nc \lim_{\epsilon \to 0}
   \Bigl[ \mathrm{Im}~
   {\tilde n}_q ( \theta=\qrw^\mp )\Bigr],
\label{Eq:QNHaTrw}
\end{align}
where
$\qrw^\mp = \theta_\mathrm{RW} \mp \epsilon =\pi / \nc \mp \epsilon$
with the positive infinitesimal value $\epsilon$.
Below $T_\mathrm{RW}$,
$n_q(\theta=\pi/\nc)$ is exactly zero
and thus $\Psi=0$, but $\Psi$ becomes non-zero above $T_\mathrm{RW}$.
The coefficient $\nc$ in Eq.~(\ref{Eq:QNHaTrw}) reflects the number of
the gapped point in the $0 \le \theta \le 2 \pi$ region.

In the case B where the RW endpoint is the triple point,
situations become complicated in the
$T_\mathrm{Beard} < T < T_\mathrm{RW}$ region.
In this region, ${\tilde n}_q$ jumps from high-$T$ curve to the low-$T$
curve at $\theta=\theta_\mathrm{B}$ where $\theta_\mathrm{B}$ is defined
as the dimensionless imaginary chemical potential on the beard line.
The quark number holonomy can be expressed as
\begin{align}
\Psi
&= 2 \nc \lim_{\epsilon \to 0}
   \Bigl[ \mathrm{Im}
   \Bigl\{
          {\tilde n}_q \Bigl( \theta=\theta_\mathrm{B}^-
                       \Bigr)
        - {\tilde n}_q \Bigl( \theta=\theta_\mathrm{B}^+
                       \Bigr)
   \Bigr\} \Bigr]
\nonumber\\
& \neq 0,
\label{Eq:QNHbTrw}
\end{align}
where $\theta_\mathrm{B}^\mp$ mean $\theta_\mathrm{B} \mp \epsilon$.
The number of the gapped point is $2 \nc$ in the $0 \le \theta \le 2
\pi$ region above $T_\mathrm{Beard}$.
Below $T_\mathrm{Beard}$, $\Psi$ should be zero.

\subsection{Asymptotic behavior}
\label{Sec:AB}

Here, we discuss the asymptotic behavior of $\Psi$.
The quark number holonomy in the $T \to \infty$ limit becomes
\begin{align}
\Psi^{T \to \infty}
&= 2 \nc \lim_{\epsilon \to 0}
   \Bigl[ \mathrm{Im}~{\tilde n}_{q}^{T \to \infty}
   ( \theta = \qrw^- ) \Bigr]
\neq 0,
\end{align}
where ${\tilde n}_{q}^{T \to \infty}$ is the normalized quark number
density in the $T \to \infty$ limit.
Actual value of ${\tilde n}_q^{T \to \infty}$ can be obtained from the
perturbative one-loop effective
potential~\cite{Gross:1981br,Weiss:1980rj,Kashiwa:2013rm} and the value
becomes
\begin{align}
\lim_{\epsilon \to 0}
\frac{\mathrm{Im}~{\tilde n}_q^{T \to \infty} ( \theta = \qrw^\mp )}
     {\nc \nf}
& = \pm \frac{2 \pi}{3}
    \Bigl[ \frac{\qrw}{2\pi}
         - 4 \Bigl(\frac{\qrw}{2\pi}\Bigr)^3
    \Bigr]
\nonumber\\
& \xrightarrow[\nc = 3]{} \pm 0.31 \cdots,
\end{align}
where the normalization constant $C$ is set to $T^{-3}$.

The quark number holonomy in the $m \to \infty$ limit can be discussed
by using the hopping parameter expansion in the lattice formalism.
From the straightforward calculation,
the normalized quark number density is obtained as
\begin{align}
&\lim_{\epsilon \to 0}
\Bigl[
\mathrm{Im}~{\tilde n}_q (\theta = \qrw^-)
\Bigr]
\nonumber\\
& \hspace{10mm}
 = - \frac{C \nf}{N^3_s N_t}
     \lim_{\epsilon \to 0}
     \sum_{n=1}^\infty
     \kappa^n
     \mathrm{Im}
     \Big\langle \mathrm{Tr}
     \Bigl(
     \frac{\partial Q}{\partial \mu} Q^{n-1}
     \Bigr)
     \Big\rangle,
\label{Eq:hpe}
\end{align}
where the hopping parameter $\kappa$ is related to $m$ as
$\kappa = 1/(2m+8)$, $N^3_s N_t$ express the space-time lattice volume,
$\langle \cdots \rangle$ means the configuration average and $Q$ is the
covariant derivative part of the lattice action; see
Ref.~\cite{Aarts:2014bwa} as an example.
Since the hopping parameter is $\kappa \propto 1/m$ in the heavy quark
mass limit, the quark number holonomy is suppressed by $1/m$.
Thus, the quark number holonomy finally becomes zero at $m=\infty$.
However, the confinement-deconfinement transition temperature determined
by the quark number holonomy perfectly matches with the transition
temperature determined by $\Phi$ at $m=\infty$ by carefully considering
$m \to \infty$ limit.

Even in the heavy quark mass region, the RW transition may be smeared by
the finite size effect in the lattice QCD simulation.
Thus, the non-zero quark number holonomy requires
finite size scaling analysis to obtain in a straightforward calculation.
An alternative way is to fit lattice QCD data by using an
oscillating $\theta$-odd functions in the region
$V$, $\{ V : 0 \le \theta \le \pi/\nc \}$,
with neglecting data very close to $\theta = (2k-1)\pi/\nc$
where data are strongly affected by the finite size effects.
The fitting function becomes the $2\pi/\nc$ periodic
function at sufficiently low $T$ and it does the $2 \pi$ periodic
function at sufficiently high $T$ in the $0 \le \theta \le \pi / \nc$
region.
This difference may help us to calculate the quark number holonomy on
the lattice.
It should be noted that this treatment is similar to the observation
process of non-zero order parameters with vanishing symmetry breaking
external fields on the lattice and thus it is not a fundamental problem.

\section{Discussions}
\label{Sec:Discussion}

Firstly, we discuss the current status of the present determination and
the ordinary determination of the deconfinement temperature.
Readers may doubt the validity of the present definition of the
deconfinement transition temperature since the deconfinement
temperature, $T_\mathrm{D} \equiv T_\mathrm{RW}$ or $T_\mathrm{Beard}$,
is substantially higher than the chiral pseudo-critical temperature.
It was considered that the chiral and the deconfinement crossover take
place at similar temperatures from the rapid change of the
chiral condensate and the Polyakov-loop on the
lattice with $2+1$ flavors; see for example
Ref.~\cite{Aoki:2006br,Cheng:2007jq}.
With the development of the highly improved quark action, it now
seems that the Polyakov-loop grows very
gradually~\cite{Borsanyi:2010bp,Bazavov:2011nk}.
An effective model analysis of recent lattice data implies that
the deconfinement pseudo-critical temperature ($\sim 215$ MeV) is
substantially higher
than the chiral pseudo-critical
temperature~\cite{Miyahara:2016din}.
By comparison, a recent lattice determination of the RW endpoint
temperature with physical quark masses implies that
the continuum extrapolated value of $T_\mathrm{RW}$ is
$208(5)$ MeV~\cite{Bonati:2016pwz}.
Therefore, higher $T_\mathrm{D}$ does not invalidate the
discussion, but is supported by the recent lattice data via effective
model analysis.

Secondly, we discuss the difference between the quark number holonomy and the dual
quark
condensate~\cite{Bilgici:2008qy,Bilgici:2009tx,Bilgici:2009phd,Fischer:2009wc,Kashiwa:2009ki}.
The dual quark condensate is defined as
\begin{align}
\Sigma^{(n)}
&= - \oint_{0}^{2\pi} \frac{d\varphi}{2\pi} e^{-in\varphi} \sigma(\varphi),
\label{Eq:DCC}
\end{align}
where $\varphi=\theta+\pi$
specifies the boundary condition for the
temporal direction of quarks, $\sigma(\varphi)$ is the
$\varphi$-dependent chiral condensate and $n$ represents the winding
number along the temporal direction.
Particularly, $\Sigma^{(1)}$ shares similar properties with $\Phi$
because $\Phi$ is also the winding number $1$ quantity and thus
it can be used as the indicator of the confinement-deconfinement
transition.
In the quenched approximation, the dual quark condensate is well
defined, but there is the uncertainty in the dynamical
quark case~\cite{Bilgici:2009tx,Bilgici:2009phd}.
In the calculation of the dual quark condensate, we
need to break the RW periodicity because $\Sigma^{(1)}$ should be zero in
all $T$ region if the RW periodicity exists.
It is usually done by imposing the twisted boundary condition on the
Dirac operator, while configurations are sampled under the anti-periodic
boundary condition.
This is not a unique procedure.
Therefore, there is the uncertainty in the determination of the dual
quark condensate.
Also, it is well known that the dual quark condensate is strongly
affected by the chiral transition or some other transitions
\cite{Benic:2013zaa,Marquez:2015bca,Zhang:2015baa}.
On the other hand, the quark number holonomy (\ref{Eq:holonomy_n}) can
provide non-zero value above $\trw$ or $T_\mathrm{Beard}$ without
any uncertainties.
It is the important advantage of the quark number holonomy.

Thirdly, we discuss the quark number holonomy from the landscape of the
effective potential in the complex $\Phi$ plane.
Below $\trw$ or $T_\mathrm{Beard}$, the effective potential at any
$\theta$ can be described by only one minimum which is continuously
connected with the $\theta=0$ solution.
Above $\trw$, the $\zc$ images appear;
for example, the $\zc$ images are $e^{i2\pi/3}$ and $e^{i 4\pi/3}$ for
$\Phi=1$ at sufficiently high $T$ for $\nc=3$.
In the confined phase, the fluctuation is strong and thus the $\zc$
images are collapsed to one minimum.
On the other hand, the $\zc$ images can withstand the fluctuation in the
deconfined phase.
Thus, the quark number holonomy (\ref{Eq:holonomy_n})
measures the strength of the fluctuation
which collapses the $\zc$ images to one minimum.
It is related to the non-trivial degeneracy of the free-energy in
the deconfined phase discussed in Ref.~\cite{Kashiwa:2015tna}.
Therefore, the quark number holonomy can describe the
confinement-deconfinement transition via the nontrivial free-energy
degeneracy.
Present discussion may be related with the Polyakov-loop fluctuations
discussed in Ref.~\cite{Lo:2013hla} and thus it is interesting to
compare the results of the Polyakov-loop fluctuations with the quark
number holonomy.

Finally, the sign problem is discussed when we calculate the quark
number holonomy at finite $\mu_\mathrm{R}$.
At finite $\mur$, Eq.(\ref{Eq:holonomy_n}) should be replaced as
\begin{align}
\Psi(T) \to \Psi(T,\mur).
\label{Eq:holonomy_n_R}
\end{align}
This means that the $\theta$ integration in Eq.~(\ref{Eq:holonomy_n_R})
should be evaluated with fixed $T$ and also $\mur$.
Therefore, we must consider the complex chemical potential in the
calculation of the quark number holonomy, where the sign problem arises.
At finite imaginary chemical potential ($\mur=0$),
we can use the $\gamma_5$ hermiticity;
\begin{align}
   \mathrm{det} {\cal D}(\mu)
 = \mathrm{det} [\gamma_5 {\cal D} (\mu) \gamma_5]
 = [\mathrm{det} {\cal D}(-\mu^*)]^*,
\label{Eq:zero_mu}
\end{align}
where ${\cal D}$ is the Dirac operator.
Therefore, the sign problem does not matter at finite
imaginary chemical potential, $\mu^* = - \mu$
, when we calculate $\Psi(T)$.
On the other hand, at finite real chemical potential,
$\mur \neq 0$ and $\theta=0$,
the relation (\ref{Eq:zero_mu}) can not help us, but the
Lefschetz thimble path integral
method~\cite{Witten:2010cx,Cristoforetti:2012su,Fujii:2013sra} does.
In Ref.~\cite{Tanizaki:2015pua}, it is shown that this method leads
the saddle-points which manifests the ${\cal C} {\cal K}$
symmetry where ${\cal C}$ and ${\cal K}$ express the charge and the
complex conjugation operator, respectively.
The sign problem can be avoided by the ${\cal C} {\cal K}$
symmetric saddle-points and then the mean-field calculation
of QCD effective models such as the PNJL model is extremely
simplified~\cite{Nishimura:2014rxa,Nishimura:2014kla}.
Unfortunately, the ${\cal C} {\cal K}$ symmetry
is not preserved at finite complex chemical potential,
$\mur \neq 0$ and $\theta\neq 0$,
when we calculate Eq.~(\ref{Eq:holonomy_n_R}).
Thus, the calculation becomes complicated even in the
mean-field calculation of the QCD effective models.
In this case, we should perform the matter-of-fact calculation based on
the Lefschetz thimble path integral method.
Actual challenge of the calculation will be shown elsewhere.

\section{Summary}
\label{Sec:Summary}

In this paper, we have proposed a new quantity to describe the
confinement-deconfinement transition based on topological properties
of QCD in the imaginary chemical potential region.
We call it the quark number holonomy which is defined by the contour
integral of the quark number susceptibility along the closed loop
of $\theta$.
The quark number holonomy seems to be similar to
the Uhlmann phase which can be used to classify the topological order at
finite $T$ in the condensed matter physics.

The quark number holonomy can have a non-zero value above
$T_\mathrm{RW}$ or $T_\mathrm{Beard}$ and it becomes zero below
these temperatures.
This behavior is related with the different realizations of the
free-energy degeneracy above and below $T_\mathrm{RW}$.
From the model independent analysis,
we find that the quark number holonomy is proportional to $\nc^2$
in the deconfined phase, while it does not in the confined phase if we
determined the confinement-deconfinement temperature as the topological
phase transition.
Also, we have shown the behavior of the quark number holonomy in the
$T\to\infty$ and the $m\to\infty$ limit as a landmark to help the future
lattice QCD simulation.

We have discussed the similarity between the quark number holonomy and
the dual quark condensate which is sometimes used to investigate the
confinement-deconfinement transition.
Calculations of the dual quark condensate has the uncertainty when the
dynamical quark is taken into account, but the quark number holonomy
does not have such uncertainty.
This is the strong advantage of the quark number holonomy.
Also, we have explained how the quark number holonomy can describe
the confinement-deconfinement transition from the landscape of the
effective potential.
In the confined phase, the fluctuation is strong and then the $\zc$
images collapse to one minimum.
On the other hand, the $\zc$ images withstand against the fluctuation
in the deconfined phase.
Therefore, the quark number holonomy can describe the
confinement-deconfinement transition via the free-energy degeneracy.
Finally, we have discussed the sign problem when we calculate the quark
number holonomy at finite real chemical potential.
In this case, we should consider the complex chemical potential and thus
we need extremely care of the sign problem.

\begin{acknowledgments}
The authors thank M. Sato whose suggestions for the quantum order
 parameter of the topological phase transition at finite temperature
 motivate us to start this study.
K.K. thanks H. Tsukiji for helpful comments.
K.K. is supported by Grants-in-Aid for Japan Society for the Promotion
 of Science (JSPS) fellows No.26-1717.
A.O. is supported in part by the Grants-in-Aid for Scientific Research
 from JSPS (Nos. 15K05079, 15H03663), the Grants-in-Aid for Scientific
 Research on Innovative Areas from MEXT (No. 2404: 24105001, 24105008),
 and by the Yukawa International Program for Quark-Hadron Sciences.
\end{acknowledgments}

% Appendixes %
\bibliography{ref.bib}

\end{document}